\documentclass{ckm}                 

\usepackage{txfonts}            

\confname{Workshop on the CKM Unitarity Triangle, IPPP Durham, April
  2003}

\title{Determining weak phases using $B \to D^*V$ Decays\thanks{Talk presented by Nita Sinha}}

\author{David London\addressmark{a}, Nita Sinha\addressmark{b} and Rahul Sinha\addressmark{b}}

\address[a]{ Laboratoire Ren\'e J.-A. L\'evesque, 
Universit\'e de Montr\'eal,\\
C.P. 6128, succ. centre-ville, Montr\'eal, QC,
Canada H3C 3J7}
\address[b]{Institute of Mathematical Sciences, Taramani,
 Chennai 600113, India}

\def\beq{\begin{equation}}
\def\eeq{\end{equation}}
\def\bea{\begin{eqnarray}}
\def\eea{\end{eqnarray}}
\def\nn{\nonumber}
\def\sss{\scriptscriptstyle}

\def\bd{B_d^0}
\def\bdbar{{\overline{B_d^0}}}
\def\bs{B_s^0}
\def\bsbar{{\overline{B_s^0}}}
\def\bbar{{\overline{B^0}}}
\def\barp{{\raise.35ex\hbox
{${\sss (}$}}---{\raise.35ex\hbox{${\sss )}$}}}
\def\bdbarp{\hbox{$B_d$\kern-1.4em\raise1.4ex\hbox{\barp}}}
\def\bsbarp{\hbox{$B_s$\kern-1.4em\raise1.4ex\hbox{\barp}}}
\def\barpd{{\raise.35ex\hbox
{${\sss (}$}}--{\raise.35ex\hbox{${\sss )}$}}}
\def\dbarp{\hbox{$D^{*0}$\kern-1.6em\raise1.5ex\hbox{\barpd}}}
\def\kbarp{\hbox{$K^{*0}$\kern-1.6em\raise1.5ex\hbox{\barpd}}}
\def\dpbarp{\hbox{$D^{0}$\kern-1.2em\raise1.5ex\hbox{\barpd}}}
\def\ks{K_{\sss S}}

\def\roughly#1{\mathrel{\raise.3ex\hbox
{$#1$\kern-.75em\lower1ex\hbox{$\sim$}}}}

\def\adir00{{a_{\sss dir}^{00}}}

\def\B00{B^{00}}
\def\Bp0{B^{+0}}

\def\epjc#1#2#3{{\it Eur.\ Phys.\ J.}\ {\bf C#1}, #3 (19#2)}
\def\epjcn#1#2#3{{\it Eur.\ Phys.\ J.}\ {\bf C#1}, #3 (20#2)}

\def\npb#1#2#3{{\it Nucl.\ Phys.} {\bf B#1}, #3 (19#2)}
\def\plb#1#2#3{{\it Phys.\ Lett.} {\bf #1B}, #3 (19#2)}
\def\prd#1#2#3{{\it Phys.\ Rev.} {\bf D#1}, #3 (19#2)}

\def\prl#1#2#3{{\it Phys.\ Rev.\ Lett.} {\bf #1}, #3 (19#2)}
\def\zpc#1#2#3{{\it Zeit.\ Phys.} {\bf C#1}, #3 (19#2)}

\begin{document}

\begin{abstract}
We describe how the angular analysis of vector-vector final states in
  $B$ decays provides theoretically clean techniques for determination
  of CP violating phases. The quantity $\sin^2 (2\beta + \gamma)$ can
  be cleanly obtained from the time dependent study of decays such as
  $\bd(t) \to D^{*\pm} \rho^\mp$, $D^{*\pm} a_1^{\mp}$ etc.
  Similarly, one can use $\bs(t) \to D_s^{*\pm} K^{*\mp}$ to extract
  $\sin^2 \gamma$. A time independent study of the charged decay modes
  $B^\pm\to \dbarp~~ K^{*\pm}$ can also be used to extract $\gamma$.
\end{abstract}

\maketitle


\section{Why look at VV modes for determining CPV Phases?}
\label{first}

A precise determination of all CP violating angles~\cite{CPreview} is
one of the major goals of the current and future B Physics
experiments. The measured values of these angles may be consistent
with the standard model (SM) predictions, or they may indicate the
presence of physics beyond the standard model. Early indications are
that new physics, if present, is likely to have rather small effects
on the angles of the unitarity triangle. Hence, to uncover any new
physics, it is extremely important that all the CP violating angles be
determined without theoretical uncertainties.

 In the early days of the field, it was thought that the CP angles
could be easily measured in $\bd(t) \to \pi^+ \pi^-$ ($\alpha$),
$\bd(t) \to \Psi\ks$ ($\beta$), and $\bs(t) \to \rho\ks$
($\gamma$). However, it soon became clear that things would not be so
easy: the presence of penguin amplitudes~\cite{penguins} makes the
extraction of $\alpha$ from $\bd(t) \to \pi^+ \pi^-$ quite difficult,
and completely spoil the measurement of $\gamma$ in $\bs(t) \to
\rho\ks$. In fact, determining $\gamma$ through techniques which are
theoretically clean as well as experimentally feasible has been a
challenge.

The problem of penguin pollution can be avoided by considering
decay modes that involve only tree amplitudes.
 In this talk we first indicate the practical problems encountered by
most of the clean methods to determine $\gamma$, using
pseudoscalar-pseudoscalar (PP) and pseudoscalar-vector (PV) final
states and involving only tree amplitudes. We next demonstrate, how
the use of the corresponding vector-vector (VV) final state modes,
resolves these problems. The rich kinematics of the VV modes
accessible via an angular analysis, provides a large number of
observables which allows clean extraction of CP phases. 

As a first example, we consider the PP or PV final states, $f\equiv
D^-\pi^+ (D^{* -}\pi^+)$. Both $B^0$ and $\bbar$ can decay to
these final states and only one weak amplitude contributes. Hence, the
amplitudes for these decay modes may be written as,
\label{amp1}
\bea 
A\equiv Amp (B^0 \to f) &=& a e^{i\delta^a} e^{i\phi_a} ~, \\
A'\equiv Amp (\bbar \to f) &=& b e^{i\delta^b} e^{i\phi_b} ~. 
\eea
where, $\phi_a=0$ and $\phi_b=-\gamma$. Because of $B^0$--$\bbar$
mixing, CP violation comes about due to an interference between the
amplitudes $B^0 \to f$ and $B^0 \to \bbar \to f$. Note that since both
$B^0$ and $\bbar$ can also decay to ${\bar f}$, one can measure the
four time dependent decay rates, $\Gamma(\bd(t) \to f)$,
~$\Gamma(\bdbar \to f)$, ~$\Gamma(\bd(t) \to \bar{f})$ and
$\Gamma(\bdbar \to \bar{f})$. It is therefore possible to determine
the weak phase $\phi=(2\beta+\gamma)$~\cite{BDpi,ADKL}. However, the
decay amplitude, $b<<a$ and hence, the decay rate for $\Gamma(\bdbar
\to D^-\pi^+)$ is expected to be small. The ratio, $\Gamma(\bdbar \to
D^- \pi^+) / \Gamma(\bd \to D^- \pi^+)$, is essentially $|V_{ub}
V_{cd}^* / V_{cb}^* V_{ud}|^2 \simeq 4 \times 10^{-4}$. Obviously, it
will be very difficult to measure this tiny quantity with any
precision, and therefore it would not be viable to carry out this
method in practice. On the other hand, we will demonstrate that with
the corresponding VV final states ($D^{*\pm} \rho^\mp$), one can
extract $\sin^2 (2\beta + \gamma)$ using an angular analysis, without
the knowledge of the smaller amplitude~\cite{LSS2}.

The second example that we examine, is that of direct CP violation, in
the modes: $B^{\pm} \to \dpbarp~~ K^\pm$, $B^{\pm} \to \dpbarp~~
\pi^\pm$.  In this method~\cite{GroWyler}, $\gamma$ is obtained from
an interference of the mode $B\to D^0K$ with $B\to\overline{D^0}K$,
which occurs if and only if, both $D^0$ and $\overline{D^0}$ decay to
a common final state $f$; in particular, $f$ is taken to be a $CP$
eigenstate. This technique of extracting $\gamma$ requires a
measurement of the branching ratio for $B^+\to D^0 K^+$ which is not
experimentally feasible as pointed out in~\cite{ADS}. Moreover, the
$CP$ violating asymmetries tend to be small as the interfering
amplitudes are not comparable. The use of non--$CP$ eigenstates `$f$'
has also been considered~\cite{Dunietz-GLW} in literature. Atwood,
Dunietz and Soni (ADS)~\cite{ADS} extended this proposal by
considering `$f$' to be non--$CP$ eigenstates that are also doubly
Cabbibo suppressed modes of $D$. The two interfering amplitudes then
are of the same magnitude resulting in large asymmetries. Their
proposal is to use two final states $f_1$ and $f_2$ with at least one
being a non--$CP$ eigenstate. The use of more than one final state
enables not only the determination of $\gamma$, but also of all the
strong phases involved and the difficult to measure branching ratio
$Br(B^+\to D^0 K^+)$. However, an input into the determination of
$\gamma$ is the branching ratio of the doubly Cabbibo suppressed mode
of $D$. Here again, the VV final states, provide an
alternative~\cite{flambda}. (Some other plausible methods have also
been recently proposed~\cite{recent}.) The VV modes
$\dbarp~~~K^{*\pm}$, $\dbarp~~\rho^{\pm}$, enable extraction of
$\gamma$ and all the unknowns involved, including the BR for Doubly
Cabibbo-suppressed mode of $D$.

\section{Vector-Vector final state decay amplitudes}

The most general covariant amplitude for a $B$ meson decaying to a
pair of vector mesons has the form~\cite{Valencia}
\begin{eqnarray} 
 A(
\displaystyle
 B(p)\to V_1(k,\epsilon_1) V_2(q,\epsilon_2))= 
\displaystyle
\epsilon_1^{*\mu}\epsilon_2^{*\nu} \times \nn \\[2ex] 
\left(
 a\; g_{\mu\nu}+ 
\frac{b}{m_1 m_2} p_{\mu} p_{\nu} 
+i\frac{c}{m_1 m_2} 
\epsilon_{\mu\nu\alpha\beta} k^\alpha q^\beta 
\right)~,
\end{eqnarray} 
where, $\epsilon_1$, $\epsilon_2$ and $m_1$, $m_2$ 
represent the polarization vectors and the masses of the vector mesons 
$V_1$ and $V_2$ respectively. The
coefficients $a$, $b$, and $c$ can be expressed in terms of the linear
polarization basis $A_{\|}$, $A_\perp$ and $A_0$ as follows:
\begin{eqnarray}
A_0&=&- x a-(x^2-1) b ~, \nn \\
A_\|&=&\sqrt{2}a ~, \\
A_\perp&=&\sqrt{2(x^2-1)}\,c ~, \nn
\end{eqnarray}
where $x=k.q/(m_1 m_2)$.  If both mesons subsequently decay into two
$J^{P}\!=\!0^{-}$ mesons, {\it i.e.\/} $V_1\!\to\! P_1 P_1^\prime$ and
$V_2 \to P_2 P_2^\prime$, the amplitude can be expressed
as~\cite{flambda,flamdun}
\begin{eqnarray} 
A(\displaystyle B\to V_1 V_2)\propto (A^0 
\cos\theta_1 \cos\theta_2 + \frac{A^\|}{\sqrt{2}} \nonumber \\
\sin\theta_1 \sin\theta_2 \cos\phi 
-i \frac{A^\perp}{\sqrt{2} }\sin\theta_1
\sin\theta_2 \sin\phi), 
\label{ang-dist}
\end{eqnarray}
where $\theta_1$($\theta_2$) is the angle between the $P_1$($P_2$)
three-momentum vector, $\vec{k_1}(\vec{q_1})$ in the $V_1 ( V_2)$ rest
frame and the direction of total $V_1$ ($V_2$) three-momentum vector
defined in the $B$ rest frame.  $\phi$ is the angle between the
normals to the planes defined by $P_1 P_1^\prime$ and $P_2
P_2^\prime$, in the $B$ rest frame.

\section{Time dependent analysis in $B \to VV$}

We consider a final state $f$, consisting of two vector mesons, to
which both $B^0$ and $\bbar$ can decay. If only
one weak amplitude contributes to $B^0 \to f$ and $\bbar \to f$, we
can write the helicity amplitudes as follows:
\begin{eqnarray}
\label{amp1}
A_\lambda \equiv Amp (B^0 \to f)_\lambda &=& 
       a_\lambda e^{i\delta_\lambda^a} e^{i\phi_a} ~, \\ 
A'_\lambda \equiv Amp (\bbar \to f)_\lambda &=& 
       b_\lambda e^{i\delta_\lambda^b} e^{i\phi_b} ~, \\ 
{\bar A}'_\lambda \equiv Amp (B^0 \to {\bar f})_\lambda &=& 
       b_\lambda e^{i\delta_\lambda^b} e^{-i\phi_b} ~, \\ 
{\bar A}_\lambda \equiv Amp (\bbar \to {\bar f})_\lambda &=& 
       a_\lambda e^{i\delta_\lambda^a} e^{-i\phi_a} ~,
\label{amp4}
\end{eqnarray}
where the helicity index $\lambda$ takes the values $\left\{
  0,\|,\perp \right\}$. In the above, $\phi_{a,b}$ and
$\delta^{a,b}_\lambda$ are the weak and strong phases, respectively.
Using CPT invariance, the total decay amplitudes can be written as
\begin{eqnarray}
{\cal A} = Amp (B^0\to f) = A_0 g_0 + A_\| g_\| + i \, A_\perp g_\perp~~ ,
\label{A}\\
{\bar{\cal A}} = Amp (\bbar\to {\bar f}) = 
     {\bar A}_0 g_0 + {\bar A}_\| g_\| - i \, {\bar A}_\perp g_\perp~~ ,
\label{Abar}\\
{\cal A}' = Amp (\bbar\to f) = A'_0 g_0 + A'_\| g_\| - i \, A'_\perp g_\perp~~ ,
\label{Aprime}\\
{\bar{\cal A}}' = Amp (B^0 \to {\bar f}) = 
   {\bar A}'_0 g_0 + {\bar A}'_\| g_\| + i \, {\bar A}'_\perp g_\perp~~ ,
\label{Aprimebar}
\end{eqnarray}
where the $g_\lambda$ are the coefficients of the helicity amplitudes,
defined using Eq.~(\ref{ang-dist}) and depend only on the angles
describing the kinematics.
With the above equations, the time-dependent decay rate for a $B^0$
decaying into the two vector--meson final state, {\em i.e.\/} $B^0(t)
\to f$, can be written as
\bea
\Gamma(B^0(t) \to f)= e^{-\Gamma t} \sum_{\lambda\leq\sigma}
~~\Bigl(\Lambda_{\lambda\sigma} + \Sigma_{\lambda\sigma}\cos(\Delta M t)\nn\\
~~~~~~~~~~~~~~~~~~~~~~~~~~~- \!\rho_{\lambda\sigma}\sin(\Delta M t)\Bigr)~ g_\lambda g_\sigma\!~.
\eea
By performing a time-dependent study and angular analysis of the decay
$B^0(t)\to\! f$, one can measure the 18 observables
$\Lambda_{\lambda\sigma}$, $\Sigma_{\lambda\sigma}$ and
$\rho_{\lambda\sigma}$. In terms of the helicity amplitudes
$A_0,A_\|,A_\perp$, these can be expressed as:
\bea
&\Lambda_{\lambda\lambda}=\displaystyle
\frac{|A_\lambda|^2+|A'_\lambda|^2}{2},&
\Sigma_{\lambda\lambda}=\displaystyle
\frac{|A_\lambda|^2-|A'_\lambda|^2}{2},\nn \\[1.5ex]
&\Lambda_{\perp i}= -\!{\rm Im}({ A}_\perp { A}_i^* \!-\! A'_\perp {A'_i}^* ),
&\Lambda_{\| 0}= {\rm Re}(A_\| A_0^*\! +\! A'_\| {A'_0}^* ),
\nn \\[1.5ex]
&\Sigma_{\perp i}= -\!{\rm Im}(A_\perp A_i^*\! +\! A'_\perp {A'_i}^* ),
&\Sigma_{\| 0}= {\rm Re}(A_\| A_0^*\!-\! A'_\| {A'_0}^* ),\nn\\[1.5ex]
&\rho_{\perp i}\!=\!-\!{\rm Re}\!\Bigl(\!\frac{q}{p}
\![A_\perp^*  A'_i\! +\! A_i^* A'_\perp\!]\!\Bigr),
&\rho_{{\sss \perp \perp}}\!=\! -\! {\rm Im}\Bigl(\frac{q}{p}\,
A_\perp^* A'_\perp\Bigr),\nn\\[1.5ex]
&\rho_{\| 0}\!=\!{\rm Im}\!\Bigl(\frac{q}{p}\!
[A_\|^* A'_0\! + \!A_0^* A'_\|\!]\!\Bigr),
&\rho_{ii}\!=\!{\rm Im}\!\Bigl(\frac{q}{p} A_i^* A'_i\Bigr),
\label{defs}
\eea
where $i=\{0,\|\}$. In the above, $q/p = \exp({-2\,i\phi_{\sss M}})$,
where $\phi_{\sss M}$ is the weak phase present in $B^0$--$\bbar$
mixing.

Similarly, the decay rate for $B^0(t) \to {\bar f}$ is given by
\bea \Gamma(B^0(t)\to {\bar f}) = e^{-\Gamma t}
\sum_{\lambda\leq\sigma}
~~\Bigl( {\bar\Lambda}_{\lambda\sigma} +
{\bar\Sigma}_{\lambda\sigma}\cos(\Delta M t) \nn\\
~~~~~~~~~~~~~~~~~~~~~~~~~~~~~~~~~~~ -
{\bar\rho}_{\lambda\sigma}\sin(\Delta M t) \Bigr)~ g_\lambda g_\sigma
~.  
\eea
The expressions for these another 18 observables,
${\bar\Lambda}_{\lambda\sigma}$, ${\bar\Sigma}_{\lambda\sigma}$ and
${\bar\rho}_{\lambda\sigma}$ are similar to those given in
Eq.~(\ref{defs}), with the replacements $A_\lambda \to {\bar
A}'_\lambda$ and $A'_\lambda \to {\bar A}_\lambda$.

Angular analysis is more powerful than previously realized. Due to
the interference between the different helicity states, there are
enough independent measurement that one can obtain weak phase
information as we now show. First, we note that
\beq
\Lambda_{\lambda\lambda}={\bar\Lambda}_{\lambda\lambda}=
\frac{(a_\lambda^2+b_\lambda^2)}{2},
~~\Sigma_{\lambda\lambda}=-{\bar\Sigma}_{\lambda\lambda}=
\frac{(a_\lambda^2-b_\lambda^2)}{2}.
\label{LamSig_eq}
\eeq
Thus, one can determine the magnitudes of the amplitudes appearing in
Eqs.~(\ref{amp1})--(\ref{amp4}), $a_\lambda^2$ and $b_\lambda^2$.
However, it must be stressed that {\em the knowledge of $b_\lambda^2$
  will not be necessary within our method}. 

Next, we have
\begin{eqnarray}
\Lambda_{\perp i}\!& = &\! -{\bar\Lambda}_{\perp i}\! =\! b_{\perp} b_i
\sin(\delta_{\perp}\!-\!\delta_i\!+\!\Delta_i)-a_\perp a_i\sin(\Delta_i), \nn \\
\Sigma_{\perp i}\!& = &\! {\bar\Sigma}_{\perp i}\! =\! -b_\perp b_i
\sin(\delta_\perp\!-\!\delta_i\!+\!\Delta_i)-a_\perp a_i\sin(\Delta_i),
\label{LS}
\end{eqnarray}
where $\Delta_i \equiv \delta_\perp^a-\delta_i^a$ and $\delta_\lambda
\equiv \delta_\lambda^b-\delta_\lambda^a$. Using Eq.~(\ref{LS}) one
can solve for $a_\perp a_i\sin\Delta_i$. We will see that this is the
only combination needed to cleanly extract weak phase information.

The coefficients of the $\sin(\Delta m t)$ term, which can be
obtained in a time-dependent study, can be written as
\beq \rho_{\lambda\lambda}\! =\! \pm a_\lambda b_\lambda
\sin(\phi\!+\!\delta_\lambda),~~ {\bar\rho}_{\lambda\lambda}\!=\!\pm
a_\lambda b_\lambda \sin(\phi\!-\!\delta_\lambda),
\label{rho_eq}
\eeq
where the sign on the right hand side is positive for $\lambda=\|,0$,
and negative for $\lambda=\perp$. In the above, we have defined the CP
phase $\phi \equiv -2\phi_{\sss M} + \phi_b - \phi_a$. These
quantities can be used to determine
\beq 2 b_\lambda\cos\delta_\lambda\! =\!
\pm\frac{\rho_{\lambda\lambda}\!+\!{\bar\rho}_{\lambda\lambda}}{a_\lambda
\sin\phi},~~ 2 b_\lambda\sin\delta_\lambda \!=\!
\pm\frac{\rho_{\lambda\lambda}\!-\!{\bar\rho}_{\lambda\lambda}}{a_\lambda
\cos\phi}.
\label{delta}
\eeq
Similarly, the terms involving interference of different helicities
are given as
\begin{eqnarray}
\rho_{\perp i}\! &=&\! -a_\perp b_i
\cos(\phi\!+\!\delta_i\!-\!\Delta_i)\!-\!  a_i b_\perp
\cos(\phi\!+\!\delta_\perp\!+\!\Delta_i) , \nn\\ {\bar\rho}_{\perp
i}\! &=&\! -a_\perp b_i \cos(\phi\!-\!\delta_i\!+\!\Delta_i) \!-\!
a_i b_\perp \cos(\phi\!-\!\delta_\perp\!-\!\Delta_i) .
\label{rhocombs}
\end{eqnarray}

Putting all the above information together, we are now in a position
to extract the weak phase $\phi$. Using Eq.~(\ref{delta}), the
expressions in Eq.~(\ref{rhocombs}) can be used to yield
\begin{eqnarray}
\!\rho_{\perp i}\!+\!{\bar\rho}_{\perp i}\! &&=
-\cot\phi\,{a_i a_\perp}\cos\Delta_i\Bigg[\frac{\rho_{i
i}+{\bar\rho}_{i i}}{a_i^2}- \frac{\rho_{\sss\perp
\perp}+{\bar\rho}_{{\sss \perp \perp}}}{a_\perp^2}\Bigg] \nn \\ 
~~~~~~~~~~~~&&-{a_i a_\perp}\sin\Delta_i\Bigg[\frac{\rho_{i i}-{\bar\rho}_{i
i}}{a_i^2}+ \frac{\rho_{{\sss \perp \perp}}-{\bar\rho}_{
{\sss \perp \perp}}}{a_\perp^2}\Bigg],\\ 
\rho_{\perp i}\!-\!{\bar\rho}_{\perp i} &&=
\tan\phi\,{a_i a_\perp}\cos\Delta_i\Bigg[\frac{\rho_{i
i}-{\bar\rho}_{i i}}{a_i^2}- \frac{\rho_{\sss\perp
\perp}-{\bar\rho}_{{\sss \perp \perp}}}{a_\perp^2}\Bigg] \nn \\ 
~~~~~~~~~~~~&&-{a_i a_\perp}\sin\Delta_i\Bigg[\frac{\rho_{i i}+{\bar\rho}_{i
i}}{a_i^2}+ \frac{\rho_{{\sss \perp \perp}}+{\bar\rho}_{
{\sss \perp \perp}}}{a_\perp^2}\Bigg].
\label{sol}
\end{eqnarray}
In the above two
equations: (i) $\rho_{\lambda\sigma}$ and ${\bar\rho}_{\lambda\sigma}$
are measured quantities, (ii) the $a_\lambda^2$ are determined from
the relations in Eq.~(\ref{LamSig_eq}), and (iii) ${a_i
  a_\perp}\sin\Delta_i$ is obtained from Eq.~(\ref{LS}). Thus, the
above two equations involve only two unknown quantities, $\tan\phi$
and ${a_i a_\perp}\cos\Delta_i$, which can easily be solved for (up to a
sign ambiguity in each of these quantities). Hence, $\tan^2\phi$
(or, equivalently, $\sin^2 \phi$) can be obtained from the angular
analysis.

Note that this method relies on the measurement of the interference
terms between different helicities. However, we do not actually
require that all three helicity components of the amplitude be used.
In fact, one can use observables involving any two of largest helicity
amplitudes. In the above description, one could have chosen `$\|\,0$'
instead of `$\perp\!\|$' or `$\perp\! 0$'.

We now turn to specific applications of this method. Consider first
the final states where, $f = \pm {\bar f}$. In this case, the
parameters of Eqs.~(\ref{amp1})--(\ref{amp4}) satisfy $a_\lambda =
b_\lambda$, $\delta_\lambda^a = \delta_\lambda^b$ (which implies that
$\delta_\lambda = 0$), and $\phi_a = -\phi_b$ (so that $\phi \equiv
-2\phi_{\sss M} + 2 \phi_b$). As described above, $a_\lambda^2$ can be
obtained from Eq.~(\ref{LamSig_eq}). But now the measurement of
$\rho_{\lambda\lambda}$ [Eq.~(\ref{rho_eq})] directly yields $\sin
\phi$. In fact, this is the conventional way~\cite{helicity} of using
the angular analysis to measure the weak phases: each helicity state
separately gives clean CP-phase information. Thus, for such states,
nothing is gained by including the interference terms.

In order to have final states with only one weak amplitude, we
consider states that do not receive penguin contributions. The only
such Cabibbo-allowed quark-level decays are ${\bar b} \to {\bar c} u
{\bar d} ,~ {\bar u} c {\bar d}$. The meson level examples of these
are $\bd/\bdbar \to D^{*-}\rho^+, D^{*+}\rho^-$. These are the VV
counterparts of the PP/PV modes described in first example of
section~\ref{first}. As we have already emphasized in the discussion
following Eq.~(\ref{LamSig_eq}), {\it none of the observables or
combinations required for the analysis to extract $\sin^2 (2\beta +
\gamma)$ are proportional to $b_\lambda^2$}. Thus, we avoid the
practical problems present in Dunietz's method~\cite{BDpi}.

The two decay amplitudes for the final states $D^{*\pm} \rho^\mp$ have
very different sizes, {\it i.e.\/} $b_\lambda \ll a_\lambda$. This
results in a very small CP-violating asymmetry whose size is
approximately $|V_{ub} V_{cd}^* / V_{cb}^* V_{ud}| \approx 2\%$.
Thankfully, the situation is alleviated by the large branching ratio
for the decay $\bd \to D^{*-}\rho^+$, roughly 1\%.

The Cabibbo-suppressed quark-level decays which do not receive penguin
contributions are ${\bar b} \to {\bar c} u {\bar s} ,~ {\bar u} c
{\bar s}$, at meson level these would correspond to $\bd \!\to \!{\bar
D}^{*0}\! K^{*0}$, $D^{*0} K^{*0}$ and $\bdbar\!\to\! D^{*0} {\bar
K}^{*0}$, ${\bar D}^{*0} {\bar K}^{*0}$, with $K^{*0}$ and ${\bar
K}^{*0}$ decaying to $\ks \pi^0$. However, in these modes one has the
old problem of taggig the neutral $D's$~\cite{ADS}.

One can also consider $\bs$ and $\bsbar$ decays.  corresponding to the
quark-level decays ${\bar b} \to {\bar c} u {\bar d} ,~ {\bar u} c
{\bar d}$, or ${\bar b} \to {\bar c} u {\bar s} ,~ {\bar u} c {\bar
  s}$.  The most promising processes are the Cabibbo-suppressed decay
modes $\bs/\bsbar \to D_s^{*\pm} K^{*\mp}$. Here the $\bs-\bsbar$
mixing phase is almost $0$, so that the quantity $\sin^2 \gamma$ can
be extracted from the angular analysis of $\bs(t)\to D_s^{*\pm}
K^{*\mp}$. Other methods for obtaining the CP phase
$\gamma$ using similar final states have also been
studied~\cite{ADK}.

\section{Direct Asymmetry in $B \to VV$}
It is {\em also possible to cleanly extract the weak phase $\gamma$
  using only charged $B^\pm$ decays,} by studying the angular
  distribution~\cite{flambda}. The decays $B^+\to D^{*0} V^+$, $B^+\to
  \overline{D^{*0}} V^+$ and $B^-\to D^{*0} V^-$, $B^-\to
  \overline{D^{*0}} V^-$ can be related by CPT. Consider
  $D^{*0}/\overline{D^{*0}}$ decaying into $D^0\pi^0 /
  \overline{D^0}\pi^0$, with $D^0/\overline{D^0}$ meson further
  decaying to a final state `$f$' that is common to both $D^0$ and
  $\overline{D^0}$. $f$ is chosen to be a Cabibbo-allowed mode of
  $D^0$ or a doubly-suppressed mode of $\overline{D^0}$. The
  amplitudes for the decays of $B^+$ and $B^-$ to a final state
  involving $f$ will be a sum of the contributions from $D^{*0}$ and
  $\overline{D^{*0}}$, and similarly for the CP-conjugate
  processes. In this case one can experimentally measure the
  magnitudes of the 12 helicity amplitudes, as well as the
  interference terms, leading to a total of 24 independent
  observables.  However, there are just 15 unknowns involved in the
  amplitudes: $a_\lambda, b_\lambda, \delta_\lambda^a,
  \delta_\lambda^b, \gamma,\Delta~{\rm and}~{\cal R}$; where, ${\cal
  R}^2=\frac{Br(\overline {D^{0}}\to f)}{Br(D^{0}\to f)},$ and
  $\Delta$ is the strong phase difference between $D^0\to f$ and
  $D^0\to {\overline f}$. Hence, the weak phase $\gamma$ may be
  cleanly extracted.

In addition, this technique has some interesting
  features:
\begin{itemize}
\item  The signals of CP violation can be obtained by adding $B$
  and $\bar{B}$ events, i.e, flavour tagging is not required.
\item  The CP
  violation signals are not diluted by the sine of strong phases.
\end{itemize}

\section{Conclusions}
We have presented new methods, that use angular analysis in $B\to VV$
decay modes, to cleanly extract weak phases. Using modes which do not
receive penguin contributions, the weak phases $(2 \beta+\gamma)$ and
$\gamma$ can be determined. We have shown that the quantity $\sin^2
(2\beta + \gamma)$ can be cleanly obtained from the time dependent
angular analysis study of the decays $\bd(t) \to D^{*\pm} \rho^\mp$,
$D^{*\pm} a_1^{\mp}$ etc.  Similarly, $\sin^2 \gamma$ can be cleanly
extracted from $\bs(t) \to D_s^{*\pm} K^{*\mp}$, or simply from an
angular analysis of the decay mode $B^\pm \to \dbarp ~~K^{*\pm}$ . The
large number of data samples already collected for the $\bd(t) \to
D^{*-} \rho^+$ mode~\cite{dstrho}, may enable, $\sin^2 (2\beta+
\gamma)$ to be the second clean measurement, after $\sin 2\beta$, to
be performed at $B$-factories.

\section*{Acknowledgements}
N.S. and R.S. would like to thank Prof. Patricia Ball and the local
organizers of the CKM Workshop for financial support. We thank the
organizers for an exciting conference. The work of D.L. was
financially supported by NSERC of Canada. The work of N.S. was
supported by the Department of Science and Technology, India.

\end{document}